# Towards a Smarter organization for a Self-servicing Society

V. De Florio
Vlaamse Instelling voor Technologisch Onderzoek (VITO)
Remote Sensing Department
2400 Mol, Belgium
vincenzo.deflorio@vito.be
University of Antwerp
2020 Antwerpen-Berchem
Belgium
vincenzo.deflorio@gmail.com

M. Bakhouya
International University of Rabat
Technopolis Rabat-Shore Rocade
Rabat-Salé, Morocco
mohamed.bakhouya@uir.ac.ma

D. Eloudghiri
Universite Moulay Ismail
Faculte des Sciences
Meknes, Morocco
dmelouad@gmail.com

C. Blondia
University of Antwerp
2020 Antwerpen-Berchem
Belgium
chris.blondia@uantwerpen.be

## ABSTRACT
Traditional social organizations such as those for the management of healthcare are the result of designs that matched well with an operational context considerably different from the one we are experiencing today. The new context reveals all the fragility of our societies. In this paper, a platform is introduced by combining social-oriented communities and complex-event processing concepts: SELFSERV. Its aim is to complement the "old recipes" with smarter forms of social organization based on the self-service paradigm and by exploring culture-specific aspects and technological challenges.

## CCS Concepts
•Applied computing: Health care information systems; •Computing methodologies: Simulation evaluation; •Human-centered computing: Collaborative and social computing systems and tools.

## Keywords
Self-service paradigm, complex event processing, service-oriented communities, health services, service-dominant logic.

## 1. INTRODUCTION
Societal and technological progress has led to a widespread of computer services — services characterized by an ever increasing complexity, pervasiveness, and social meaning. Ever more often people make use of and are surrounded by devices enabling the rapid establishment of communication means on top of which people may socialize and collaborate, supply or demand services, query and provide knowledge as it had never been possible before [7]. Assistance of the elderly population is a typical example: the share of the total population older than 65 is constantly increasing worldwide, while the current organizations still provide assistance in a non-efficient, inflexible way. New solutions for improving the health, independent living, quality of life, and active ageing of elderly citizens in the Information Society are required. We believe that one such solution could be reached by developing a collective awareness platform for socio-technical systems supporting self-servicing communities—namely, communities that actively contribute to their quality of life together with medical patients and their families.

The work presented in this paper targets the socio-technical challenge of improving the quality of life for patients (aging population, chronic diseases...) and their families. Elderly and patients require different levels of care; while in some cases hospitalization is the only answer, some of them could be treated remotely at their homes with smart close monitoring. The efficiency of this treatment would greatly benefit from the involvement of society in the whole picture. For instance, the ageing of the population, and even more the chronic diseases associated with ageing, place substantial demands on health and social care services. Traditional healthcare organizations find it difficult to meet the ensuing ever increasing demands while guaranteeing both safety and cost-effectiveness. As observed, e.g., in [25,12,13], "this requires new ways of working". In fact the "old recipes", which appeared to work in a less turbulent context, now reveal all their limitations.

As an example, diabetic monitoring is a case of healthcare service that has attracted attention in recent years. A recent study presented in the Williams textbook of endocrinology [4] states that over 382 million people suffer from diabetes throughout the world. In Morocco the situation is even worse: as declared by the president of the Moroccan League for the fight against diabetes, this disease is considered today as an epidemic in Morocco and actually "a tsunami with human, social and economic devastating [consequences]" [5]. According to data collected so far, about one and a half million people suffer from diabetes in Morocco, with a prevalence of 14% for the 50+ population class. However, almost 50% of patients are unaware of their condition since their disease is detected too late following some evident symptoms.

Furthermore, several studies, such as in [2], have shown that diabetes is one of major public health problems in Morocco, and that it causes early mortality of patients while at the same time increasing healthcare costs (cost of drugs, hospitalizations, biological tests, visits to physicians, etc.) As stated in [3], the direct cost of diabetes in Morocco is estimated to be between US $0.47 and US $1.5 billion whereas the indirect cost is estimated to be around US $2 billion.

Patients suffering from diabetes need to test frequently their blood glucose, eat healthy, perform physical activities, and possibly take medications such as insulin. These patients are, however, distressed from current healthcare services since it is based on manual monitoring and a heavy control process, which prevents most of them to follow daily control and maintain an up-to-date state of the progress of their disease. This gap in the manual monitoring and heavy control process prevents some patients to daily control and follow the progress of their disease; as a consequence, several metabolic problems have been noticed and need to be taken seriously into consideration [3,5].

As a consequence, improving the healthcare organization of services for the Moroccan population suffering of diabetes is particularly urgent and important, in particular in order to enhance the comfort of patients and reduce the treatment costs. Software tools for effective disease control are then required to allow self-monitoring of blood glucose and dietary/activities measures. This will allow filling the gap between patients and carers with the main objective to enhance comfort of patients and reduce the treatment costs. Furthermore, we plan to exploit mutualistic relationships as an important ingredient in order to improve the organization of healthcare services.

This paper analyses this issue and introduces the SELFSERV platform. SELFSERV combines patient monitoring with complex-event processing and so-called service-oriented communities [8,9,10] for continuous monitoring of patient care, with a main focus on diabetic control services (e.g., real-time monitoring of glucose, temperature, heart rate, blood pressure). SELFSERV aims to systematically identify, put on the foreground, and exploit the vast basins of "social energy" produced by our societies: viz., "the self-serve, self-organization, and self-adaptability potentials of our societies" [10].

The remainder of this paper is structured as follows. Section 2 presents a state of the art review of current practices of diabetic monitoring. In Section 3, we introduce the SELFSERV platform architecture combining service-oriented communities (SoC) and complex-event processing (CEP) concepts. Section 4 recalls some preliminary results to show the benefits of SoC and CEP. Conclusions and future work are given in Section 5.

## 2. SELFSERV RELATED CONCEPTS

Traditional healthcare organizations typically classify their players into two disjoint categories: the active users, namely professional and informal carers, and the patients and the elderly, called tertiary users and considered as incapable of any active behavior. This produces inefficiency, discrimination, and fails to promote a healthy active attitude. What is even worse, the "servicing subset," namely the portion of the social actors that are able and available to provide care to the whole set (namely, the whole population), is quickly decreasing in proportion. The problem is not merely the ever increasing social costs; rather, it is the fact that the specter of unmanageability, namely the vision of a fragile society unable to serve its citizens, is just around the corner.

A logic consequence of the above situation is the urgent need to mutate our organizational paradigms and assumptions [1]. Simply stated, we cannot afford anymore not to use the full potential of our societies. This means that the classical product-dominant logic adapted by traditional organizations operating in domains such as healthcare and crisis management cannot meet both the manageability and the resilience requirements of our organizations. New organizational design assumptions are called for, based on service-predominant logic and able to provide us with new service-oriented paradigms—in other words, new ways to perceive and manage the current state of things.

Recently, an innovative software architecture for the management of care was introduced in [6-10]—the above mentioned SoC (Service-oriented Community). SoC allows for the creation of sociotechnical systems that have been applied, e.g., to the telecare of elderly and impaired patients [7]. One such SoC was partially implemented in the software architecture for telecare that was developed by the MOSAIC group of the University of Antwerp (UA) while participating to the Flemish ICON project "Little Sister" [11] funded by iMinds and IWT (LS). The main added value of the SoC is the fact that it allows professional carers, informal carers, and the patients themselves to be orchestrated into a coherent "care ecosystem" [12,13]. In the cited references it has been shown how the SoC may be able to achieve goals including safety, e-Inclusion, non-discrimination, and cost-effectiveness, at the same time protecting society from the risk of reaching a non-manageability threshold: in other words, the risk of not being able to guarantee and sustain the care-for-all objective.

Recent results obtained through simulation indicate that the SoC outperforms traditional organizations and realizes a care ecosystem reaching the above mentioned goals [12, 13]. In particular the structural shortage of caregivers manifested by traditional care organizations can be significantly relieved, if not solved, by a full-fledged implementation of the SoC model. Results prove in particular that, the greater a society's propensity for mutual support, the larger are the returns provided by the SoC model.

Furthermore, traditional health monitoring systems use handy medical devices for recording patient data that are eventually transmitted to a remote platform for further processing. The recent advances in pervasive technologies such as wireless ad hoc networks and mobile and wearable sensor devices [22] make it possible to develop context-aware services for continuous and real-time health monitoring. These services could react to the environment changes and users' preferences with the main aim to make their life more comfortable in function of the current context (e.g. their locations, current requirements, and ongoing activities). However, handling dynamic and frequent context changes is a difficult task since real-time event/data acquisition and processing is needed. Several e-Health platforms have been recently developed for health monitoring. There platforms use sensors for health data acquisition, wireless technologies for communication, and microprocessors for data storing, filtering, and processing. As an example, the e-Health Sensor Shield [24] uses Arduino and Raspberry PI for connecting different sensors such as accelerometers and sphygmomanometers. Collected data can be used for real-time monitoring of patients and could be also used to develop other related applications and services.

It is worth noting that complex-event or information-flow processing (CEP) and stream data mining have been also proposed recently for gathering and timely analysing information (i.e., data, events) streams in order to derive conclusions from what is happening in a given situation [16, 23]. For example, CEP (e.g., ETALIS [20], CQELS [19]) could be used in many emerging applications. Inspecting credit cards transactions to prevent frauds; predicting disasters by observing the environment; preventing dangerous traffic jams by processing traffic situations; and monitoring patients' health, are all examples of applications that require information flow processing approaches. CEP's aim is to extract new knowledge (resp. predict and anticipate future situations) from continuous events (e.g., heart rate, blood pressure, temperature) that are required to generate in real-time suitable mitigation actions (resp. anticipate required actions), e.g., homecare for the elderly, or emergency response.

In this work we intend to use the e-Health Sensor Shield and combine SoC model and CEP techniques for patients monitoring in order to allow real-time gathering and efficient processing of data to generate relevant insights about ongoing situations. Our primary objective is to design a SoC addressing the needs of the Moroccan diabetic patients and hooking into the self-serving "wells" of the Moroccan societies and culture. Already ongoing joint research activities will translate in novel features specific to diabetes telecare treatment, including: adaptive monitoring (the condition of the patient will be tracked, and it will influence the frequency of communications with the doctors, activities etc.); adaptive risk threshold (determining whether the patient is in need of hospitalization or he/she may remain at home; computed as a function of the health indicators monitored through sensors and specific to the Moroccan styles of life and habits); complex event processing; stream processing of big data; and risk prediction.

Moreover, as suggested in [12, 13], SoC makes it possible to use a "cloud" of informal carers (neighbors, relatives, volunteers…) as "verificators" in order to identify false positives in a cost-effective way. Complex event processing and streaming processing allow for extracting and continuous reasoning about patients data and environment context.

As mentioned already, SELFSERV is based on the concept of social energy. More formally, there is social energy when two parties recognize a "win-win", namely the positive returns deriving from collaboration. More precisely, the concept of social energy is based on the co-creation of value carried out by two or more parties that establish a mutualistic relationship. As one can clearly see, a key prerequisite here is the ability to identify opportunities for value co-creation. The way we envision this to be done is via a custom process which we call the "Catalyst". This is exemplified in Fig. 1-2. In particular, in Fig. 1 we show two parties that are requiring services ("Request" boxes) and at the same time are able to provide services ("Provide" boxes). Figure 1 shows how the parties in a SoC transparently publish their Provides and Requests. A Catalyst subscribes to such events and is notified.

Figure 2 illustrates how every time a new Provide or Request is received, the Catalyst verifies whether mutually satisfactory Requests are available. If so, a win-win (namely, a value co-creation opportunity) is proposed to the corresponding parties. For example, a healthcare organization would normally respond to both requests by dispatching either professional or informal carers. Of course, this would imply *i)* a given social cost, to be sustained for each healthcare request; *ii)* a delay, corresponding to the time needed to identify a responder available to treat the request.

Exploiting social energy requires that the Catalyst be automatically informed of all the "Requests" and "Provides" published in a given area. A publish/subscribe method can be used to implement this. The Catalyst then identifies possible win-wins by means of a semantic matching algorithm—more information on said algorithm is available in [8,21]. As shown in Fig. 2, the Catalyst then identifies analogies in Requests and Provides such that activities of value co-creation may be proposed to two or more parties. Fig. 3 provides a more concrete example in which two persons in need of services, Mary and Ann, are informed that they may fulfill each other's Requests. No social costs need to be sustained in this case and, at least in some cases, service delivery may be initiated with little delay. In the case illustrated in Figure 3, one of the parties is Mary, an old lady who spends much of her time sitting in her sofa and watching TV. Her major Request would be to find someone to chat with. A second party is Ann, who is afraid to leave her smart home. There, she finds sanctuary from several potentially harmful healthcare conditions she suffers from, thanks to several telecare devices that constantly monitor her state. Upon receiving their Requests, a Catalyst identifies a value co-creation opportunity: by having a walk together, Mary would find the chance to talk to another person, while Ann would be able to leave her house and have a walk outside with a "living monitor" besides her.

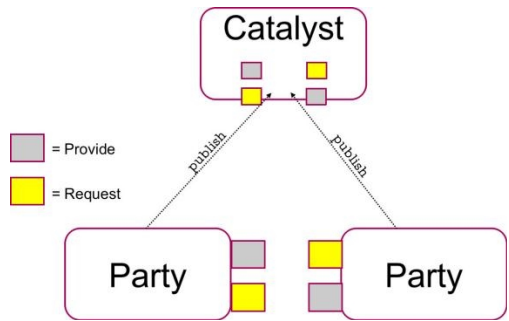

**Figure 1: The publish procedure of the parties in a SoC**

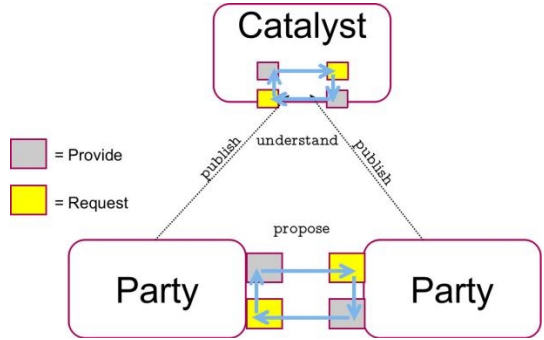

**Figure 2: The subscribe procedure of the parties in a SoC**

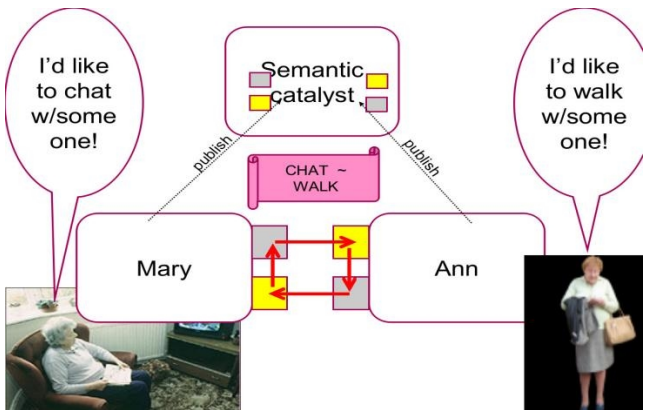

**Figure 3: A scenario of the publish/subscribe process**

## 3. SELFSERV PLATFORM

In SELFSERV we intend to demonstrate the potential of sociotechnical approaches for the organization of ambient assistance services. This will be achieved by developing a software platform for the management of an "ecosystem of care" based on telecare services. Capitalizing on the experience accrued in Flemish project "Little Sister" [11] and the Moroccan project CASANET [14, 15], SELFSERV will combine the results of these two projects into a platform able to guarantee at the same time the safety of the population suffering from diabetes and cost-effectiveness through mutualistic cooperation.

More precisely, we propose a platform that allows continuous monitoring of patient care by combining CEP and SOC techniques. This will take the shape of a web service platform structured into three levels corresponding to: *i)* monitoring people's health in their houses; *ii)* a cloud of informal carers consisting of relatives, neighbors, and volunteers; *iii)* a local hospital. In each level of the system the SELFSERV platform will "wrap" the available resources and expose them as manageable web services. The platform shall receive notifications and activate corresponding response protocols minimizing both risks and costs. Protocols will create temporary response teams that will include resources and actors from any level of the system, both professional and informal. This will be used, e.g., to reduce the societal costs associated with telemonitoring.

These concepts will be evaluated with both real experiments and simulations via two main cases studies. A first case study is to realize elements of a sociotechnical service for fall detection. Unlike purely ICT solutions, a hybrid solution in which the power of technology and the power of the people are organized into an effective sociotechnical solution is required. A second case-study shall realize elements of a sociotechnical system for telemonitoring of patients suffering from diabetes. Treatment of diabetes in Morocco is currently done by periodical checkups at hospitals. As mentioned above, this translates into discomfort and long queues for the patients and heavy workload for hospital carers. Said checkup could be reduced in number by telemonitoring the conditions of the patients and compensating for the non-optimal sensitivity and specificity through a cloud of informal carers [13].

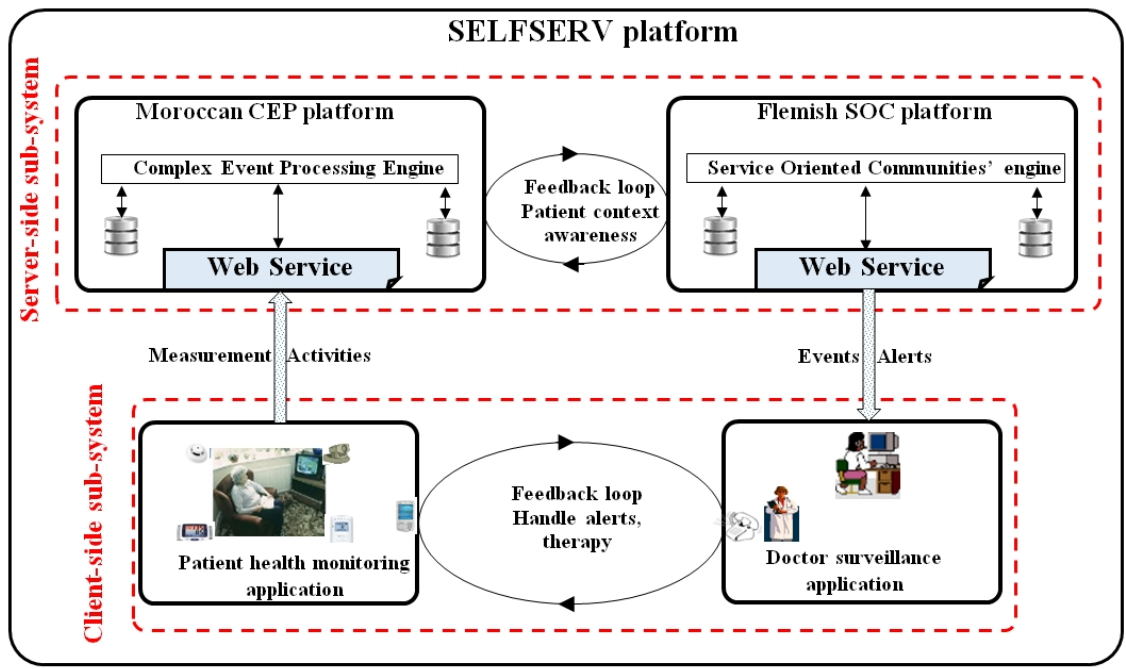

**Figure 4. The architecture of the SELFSERV platform**

As an example, let us suppose we have patients that do not usually carry out physical activities; have a bad diet; have high blood glucose and high blood pressure; and are unable to quit smoking. The doctor then decides that these patients need to be continuously monitored using the SELFSERV platform. As shown in Figure 4, each patient is equipped with on-body and wearable sensors that are connected to their smartphone via a health monitoring application [1]. All data streams are then received by the platform and processed continuously following healthcare rules, associated patterns, and patient medical incidents. All events and alerts are triggered and submitted to the doctor via the surveillance application. The doctor could then handle alerts by calling the patient on her smartphone, or informing automatically her nearby caregivers (i.e., "cloud" of informal carers) to double check her health situation, and possibly calling an ambulance for hospitalization. Using this scenario, the SELFSERV platform will be tested to show *i)* the efficiency of integrating and tailoring both CEP and SOC principles and, *ii)* its effectiveness in decreasing the waiting time at hospitals by facilitating homecare for patients (e.g., elderly) experiencing complex conditions and improving access and priorities to healthcare services as well.

## 4. PRELIMINARY RESULTS

As mentioned above, preliminary indications of the effectiveness of the SELFSERV approach were gathered by means of multi-agent simulations. In particular, in paper [13] a model of social energy was introduced and implemented with NetLogo [17] — an agent-based programming language and integrated modeling environment that is particularly well suited for modeling complex systems. Said model currently focuses on SoC and shall be augmented, in the framework of SELFSERV, with CEP elements. Our model introduces several classes of agents, representing primary, secondary, and tertiary Ambient Assisted Living users. In particular, we modeled a cloud of "verificators", namely informal carers able to move to the location of a suspected condition and report whether that condition was truly ongoing or otherwise. The model measures several metrics, including among others the social cost of each given healthcare intervention; the overall social cost sustained by the healthcare system; the number of treated cases; and the average and overall servicing time. The simulations allowed to compare the classic approach with one based on the SoC and social energy. Remarkable results were reached, including a 10% improvement in sensitivity and a drastic reduction of social costs and waiting times. A significant result was also the fact that a sensible improvement of the performance metrics are obtained with a limited number of verificators.

In parallel to this work aiming at first investigating SoC concepts and mechanisms, we are developing a platform that combines intelligent complex event processing and predictive analytics. The platform architecture includes wireless sensors network infrastructure, the CEP and stream processing engine, and data collection tool [15]. Events coming from monitored objects are continuously arriving to the platform and handled/processed by the in-memory CEP engine to provide situation-aware services and applications. It is worth noting that, despite event extraction and real-time processing of relevant information being still a challenging task, it is now possible to adapt complex-event processing techniques and use predictive analytics for analyzing streaming data in real-time in order to generate fast insights and then take actions best-matching the detected changes, thus preventing undesirable conditions from happening. For example, the increase in the glucose concentration could be predicted and the patient and doctor could be informed in advance before the concentration hits a critical level.

CEP tools are more relevant to healthcare services because they offers the ability to combine and correlate large streams of events via the use of continuous and real-time queries processing. In other words, these tools are required for developing context-awareness applications that could react to situations changes rather than simple events coming only from one source. We have then investigated in [14] three CEP engines, ETALIS [20], C-SPARQL [18] and CQELS [19] that are widely used by researchers for streaming data processing. Preliminary results using benchmark tools shows that ETALIS outperforms other engines especially in terms of scalability.

The above results are particularly encouraging; therefore, we plan to extend the SoC model by making use of ETALIS as an additional "verification layer". We expect that the combined use of social energy and intelligent sensing — namely, the SELFSERV approach — shall result in further improvements of performance and cost reductions. We are currently investigating and developing a case study related to hypoglycemia early detection using CEP techniques, mainly ETALIS, with different sensors integrated into the e-Health Sensor Shield.

## 5. CONCLUSIONS AND PERSPECTIVES

In this paper, a novel platform for the definition of diabetes telecare services is introduced. The platform combines the "self-service" model offered by the SoC platform with the advanced adaptive control features offered by the CEP techniques. This makes it possible to fast prototype and deploy personalized adaptive services able to guarantee the patient's safety with minimal impact on both the patient's quality of life and social costs. Additional simulations are ongoing. We showed also the added value of CEP techniques by recalling our preliminary results related to large events stream. Future work will focus combining both techniques into a platform for healthcare services.

## 6. ACKNOWLEDGMENTS

This work is supported by the VLIR-UOS funding agency, under the SELFSERV project (2016-2018).

## 7. REFERENCES

[1] Yalgashev O., Bakhouya, M., Nait-Sidi-Moh, A., and Gaber, J. 2016. Wireless sensor networks: basics and fundamentals. *In the book Cyber-Physical System Design with Sensor Networking Technologies*, ISBN: 978-1-84919-824-0, Eds. S. Zeadally and N. Jabeur, pp. 1-20.

[2] Bensbaa, S., Agerd, L., Boujraf, S., Araab, C., Aalouane, R., Rammouz, I., and Ajdi, F. 2014. Clinical assessment of depression and type 2 diabetes in Morocco: Economical and social components, *J. Neurosci Rural Pract*.;5(3):250-3. doi: 10.4103/0976-3147.133576.

[3] Boutayeb, W., Lamlili, M., Boutayeb, A., and Boutayeb, S. 2013. Estimation of direct and indirect cost of diabetes in Morocco, *J. Biomedical Science and Engineering*, 6, 732-738.

[4] Williams Textbook of Endocrinology, 12e Hardcover – June 10, 2011, by Shlomo Melmed MD (Author), Kenneth S. Polonsky MD (Author), P. Reed Larsen MD FRCP (Author), Henry M. Kronenberg MD (Author), ISBN-13: 978-1437703245.

[5] Belkhadir, J. 2015. *Diabetes in Morocco -- diabetes: a real tsunami*. Online; retrieved from http://www.lmlcd.com/index.php?option=com_content&view=article&id=352&Itemid=118.

[6] De Florio, V., Sun, H., Buys, J., and Blondia, C. 2013. On the Impact of Fractal Organization on the Performance of Socio-technical Systems. *Proceedings of the International Workshop on Intelligent Techniques for Ubiquitous Systems*.

[7] De Florio, V., Bakhouya, M., Coronato, A., and Di Marzo Serugendo, G. 2013. Models and Concepts for Socio-technical Complex Systems: Towards Fractal Social Organizations, *Syst. Res. Behav. Sci.* 30: 6. pp. 750-772.

[8] De Florio, V., Sun, H., and Bakhouya, M. 2014. Mutualistic Relationships in Service-Oriented Communities and Fractal Organizations. *The 2nd WCCS conference*, pp. 756-761.

[9] De Florio, V., Coronato, A., Bakhouya, M., and Di Marzo Serugendo, G. 2012. Service-oriented Communities: Models and Concepts towards Fractal Social Organizations. *The 9th SITIS Conference*, pp. 450-457.

[10] De Florio, V. and Blondia, C. 2010. Service-Oriented Communities: Visions and Contributions towards Social Organizations, On the Move to Meaningful Internet Systems. *OTM 2010 Workshops, Lecture Notes in Computer Science* Vol. 6428, pp. 319-328, Springer.

[11] Little Sister: Allowing elderly people to live autonomously for a longer period of time through the creation of a monitoring framework that uses low-resolution visual sensors. 2014. Retrieved from http://www.iminds.be/en/projects/2014/03/04/littlesister and http://www.iminds.be/~/media/iminds/images/projects/littlesister/icon%20closing%20leafletlittle%20sister%20e.ashx?la=en

[12] De Florio, V. and Pajaziti, A. 2015. How Resilient Are Our Societies? Analyses, Models, and Preliminary Results. *In Proceedings of the third WCCS*.

[13] De Florio, V. and Pajaziti, A. 2015. Tapping Into the Wells of Social Energy: A Case Study Based on Falls Identification. *Proc. of the 17th International Conference on Information Integration and Web-based Applications & Services (iiWAS)*, ACM.

[14] Lachhab, F., Bakhouya, M., Ouladsine, R., and Essaaidi, M. 2016. Performance Evaluation of CEP Engines for Stream Data Processing. *Accepted, to appear in Cloudtech*, Marrakech Morocco.

[15] Lachhab, F., Bakhouya, M., Ouladsine, R., and Essaaidi, M. 2016. Towards a Context-aware Platform for Complex and Stream Event Processing. *To appear in HPCS 2016*, Innsbruck, Austria.

[16] Cugola, G. and Margara, A. 2012. Processing flows of information: From data stream to complex event processing, *ACM Computing Surveys*, Volume 44 Issue 3.

[17] https://ccl.northwestern.edu/netlogo/

[18] Barbieri, D. F., Braga, D., Ceri, S., Della, E. V., and Grossniklaus, M. 2019. C-SPARQL: SPARQL for continuous querying. *In: Proceedings of the 18th International Conference on World Wide Web*. ACM, New York, pp 1061-1062.

[19] Le-Phuoc, D., Dao-Tran, M., Parreira, J. X., and Hauswirth, M. 2011. A native and adaptive approach for unified processing of linked streams and linked data. *In: Proceedings of the 10th ISWC—Volume Part I, ISWC'11*. Springer, Berlin, pp. 370-388..

[20] Anicic, D., Fodor, P., Rudolph, S., and Stojanovic, N. 2010. A rule-based language for complex event processing and reasoning. *In Conference on Web Reasoning and Rule Systems* (RR 2010).

[21] Sun, H., De Florio, V., and Blondia, C. 2013. Implementing a Role Based Mutual Assistance Community with Semantic Service Description and Matching. *In: Proceedings of the Inter. ACM MEDES Conference*, Luxembourg.

[22] Hester, T., Hughes, R., Sherrill, D. M., Knorr, B., Akay, M., Stein, J., and Bonato, P. 2006. Using wearable sensors to measure motor abilities following stroke. *Inter. Workshop on Wearable and Implantable Body Sensor Networks*, pp.4-8.

[23] Gama, J. and Gaber, M. 2007. *Learning from data streams, Processing Techniques in Sensor Networks*. ISBN 3540736786.

[24] Rakay, R., Visnovsky, M., Galajdova, A., and Simsik, D. 2015. Testing properties of e-health system based on arduino, *Journal of Automation and Control* 3(3), 122-126.

[25] Deloitte, Primary care: Working differently. Telecare and telehealth – a game changer for health and social care. Deloitte Centre for Health Solutions. http://www2.deloitte.com/content/dam/Deloitte/uk/Documents/life-sciences-health-care/deloitte-uk-telehealth-telecare.pdf